\documentclass[
    ,final            % use final for the camera ready runs
  ]
  {aipproc}

\layoutstyle{6x9}

\newcommand{ \rts }{$\sqrt{s_{_{\rm NN}}}$}
 
\def \auau  {Au+Au}
\def \dau   {$d$+Au}

\def \pt    {$p_T$}

\def \d0    {$D^0$}
\def \betat {$\langle \beta_T \rangle$ }
\def \gevc  {GeV/$c$}
\def \pt    {$p_T$}

\def \mbeta {$\langle \beta_T \rangle$}
\def \la    {$\Lambda + \overline{\Lambda}$ }
\def \xi    {$\Xi^- + \overline{\Xi}^+$ }
\def \om    {$\Omega^- + \overline{\Omega}^+$ }

\begin{document}

\title{Heavy-Flavor Collectivity -- Light-Flavor Thermalization at RHIC}


\author{K. Schweda
 \footnote{Present address:
        University of Heidelberg, Physikalisches Institut, \newline   
        Philosophenweg 12, 69120 Heidelberg, Germany.}}{       
address={Lawrence Berkeley National Laboratory, One Cyclotron Rd MS70R0319, Berkeley, CA 94720, USA}
}

%\thanks{Present address:
%        University of Heidelberg, Physikalisches Institut, \newline  
%        Philosophenweg 12, 69120 Heidelberg, Germany.}
%\thanks{For the full list of STAR authors and acknowledgments, see appendix 'Collaborations' of this volume.}}

\begin{abstract}
Flow measurements of multi-strange baryons 
from \auau\ collisions
at RHIC energies demonstrate that collectivity develops before hadronization, among partons.
To pin down the partonic EOS of matter produced
at RHIC, the status of thermalization in such collisions has to be addressed. 
We propose to measure collective flow of heavy-flavor quarks, e.g. charm quarks, 
as an indicator of thermalization of light flavors ($u,d,s$). 
The completion of the time of flight barrel and
the proposed upgrade with a $\mu$Vertex detector for heavy-flavor identification
in STAR are well suited for achieving these goals. 

\end{abstract}

\maketitle

%%%%%%%%%%%%%%%%%%%%%%%%%%%%%%%%%%%%%%%%%%%%
%% MAINMATTER
%%%%%%%%%%%%%%%%%%%%%%%%%%%%%%%%%%%%%%%%%%%%

\section{Introduction}
Quantum Chromo--Dynamics (QCD) is the
theory of strong interactions. Lattice calculations of QCD 
predict that at a critical temperature of $T_c\simeq 170$~MeV
a phase transition of ordinary nuclear matter 
to a deconfined state of quarks and gluons occurs~\cite{Karsch02}. Quarks and gluons
are not confined in hadrons any more; they become asymptotically free.
Under the same conditions, chiral symmetry is approximately restored 
and quark masses are reduced from their large effective values 
in hadronic matter to their small bare masses.

In ultra--relativistic nuclear collisions, a system with a temperature
larger than the critical temperature $T_c$ is expected to be created. 
The development of collectivity 
at the partonic level (among quarks and gluons) and the degree 
of thermalization are closely related to the equation of state 
of partonic matter: Re-scattering among constituents and the density profile lead to the 
development of collective flow. In case of sufficient re-scattering, 
the system might be able to reach local thermal equilibrium. 

In this paper, we show by means of flow measurements of multi-strange baryons that
at RHIC energies collectivity develops before hadronization, among partons. 
We further suggest to measure heavy-flavor (c,b) collective flow 
to probe thermalization of light quarks.  

\section{Multi-strange Hadron Flow - Partonic Collectivity}
In ultra-relativistic nuclear collisions, measured final-state transverse--momentum 
spectra can be fit within a hydrodynamically motivated approach, with
a kinetic freeze--out temperature $T_{\rm fo}$ and a mean
collective flow velocity \mbeta\ as the relevant parameters~\cite{3_2}. 
Figure~1 shows results of those fits from \auau\ collisions at \rts = 200~GeV at RHIC
in the $T_{\rm fo}$-\mbeta\ plane. Dashed and solid lines represent 1-$\sigma$ and 2-$\sigma$ 
contours, respectively. As the collisions become more and more central, the bulk of the system
dominated by the yields of $\pi, K, p$ appears to be cooler and
develops stronger collective flow, representing a strongly
interacting system expansion. 
\begin{figure}[ht]
\includegraphics[width=.45\textwidth]{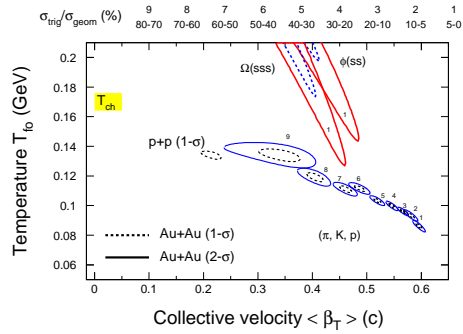}
\caption{
1--$\sigma$~(dashed lines) 
and 2--$\sigma$~(solid lines) contours for 
the transverse radial flow velocity
\betat\ and the kinetic freeze-out temperature parameter $T_{\rm fo}$ derived
from hydrodynamically motivated fits to particle spectra.
The results for $\pi, K$ and $p$, are numbered from 1~(most central) to
9~(most peripheral) \auau\ collisions and p+p
collisions~\cite{starpikp}. 
Results for the multi-strange hadrons $\phi$ and $\Omega$ are shown
in the top of for most central \auau\ collisions only.
The numbers on the top give the fraction of total hadronic cross section for centrality bins
1-9. This figure has been taken from~\cite{ActaHun}.}
\end{figure}
At the most central collisions, the
temperature parameter and the velocity are $T_{\rm fo} \sim 100$ MeV and
$\langle \beta_T \rangle \sim 0.6$(c), respectively. On the other hand, for the same
collision centrality, the multi-strange hadrons $\phi$ and $\Omega$
freeze-out at a higher temperature $T_{\rm fo} \sim 180$MeV, 
close to the point at which chemical freeze-out occurs
\cite{pbm01}. A similar behavior was also observed in \auau\ collisions at
\rts = 130 GeV \cite{xiom130}.

Multi-strange hadrons might have smaller hadronic cross sections~\cite{hsx98}
and therefore decouple from the fireball early, perhaps right at the point
of hadronization. This would explain the low \betat and higher
temperature parameter. Most importantly,
the finite value of \betat therefore must be cumulated prior to
hadronization - via partonic interactions.

Elliptic flow, due to its self-quenching nature, is an early stage signal~\cite{Sorge}.
In non-central nuclear collisions, the initial overlap
zone between the colliding nuclei is spatially deformed. If the
matter produced in the reaction zone re--scatters efficiently, this
spatial anisotropy is transferred into momentum space and the
initial, locally isotropic, momentum distribution develops
anisotropies. This anisotropy in momentum space is quantified
by the second Fourier coefficient $v_2$, the elliptic flow parameter.
%The development of anisotropies does not require
%thermalization, only re-scattering. However, 
%the largest momentum anisotropies are obtained
%in the hydrodynamic limit~\cite{1_20}, assuming zero mean free path length
%and therefore infinitely fast re--scattering which leads to instantaneous local
%thermal equilibrium distributions. 
Results on elliptic flow measurements at RHIC are shown in Fig.~2, 
(a)~for $\pi, K^0_S, p$ and \mbox{\la\ \cite{PHENIX_v2,STAR_v2KL}}, 
(b)~double-strange \xi\ \cite{STAR_v2XiO} and (c)~triple-strange \om\ \cite{STAR_v2XiO}.
The elliptic flow parameter increases with \pt\ and then saturates at larger \pt.
In the lower \pt\ region, a mass ordering is observed with lighter particles
exhibiting larger elliptic flow parameters. 
The shaded bands show results from hydrodynamical calculations for $\pi$~(upper edge)
to $\Omega$~(lower edge), 
assuming zero mean free path length
and therefore infinitely fast re--scattering which leads to instantaneous local
thermal equilibrium distributions. These calculations qualitatively describe
the experimental results in the lower \pt\ region, especially the observed mass ordering. 
 As can be seen in Fig.~2, even the multi-strange baryons \xi\ (b) and \om\ (c) 
do significantly flow.
This suggests that collective flow of multi-strange baryons \xi\ and \om\ indeed develops 
before hadronization - among partons.
\begin{figure}[hb]
  \includegraphics[height=.33\textheight]{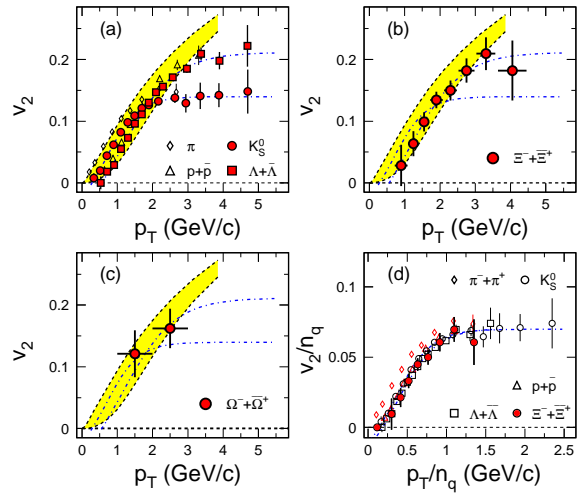}
  \caption{Results on elliptic flow measurements at RHIC for  
(a)~$\pi, K^0_S, p$ and \la~\cite{PHENIX_v2,STAR_v2KL}, 
(b)~double-strange \xi~\cite{STAR_v2XiO} and (c)~triple-strange \om~\cite{STAR_v2XiO}.
Values for $v_2$ versus \pt\ both scaled by the number $n$ of constituent quarks are shown in panel (d).
The shaded bands show results from hydrodynamical calculations for $\pi$~(upper edge)
to $\Omega$~(lower edge). The dash-dotted lines are results from empirical fit-functions~\cite{Xin04}
for baryons (upper) and mesons (lower).
}
\end{figure}

In the intermediate \pt\ region (2-6 \gevc), the calculations overshoot the data. 
At these momenta, the mean 
free path length is relatively large leading to deviations from hydrodynamic behavior.
In this region, the saturation level depends on particle type: Baryons saturate at larger
values than mesons. The dash-dotted lines are results from empirical fit-functions~\cite{Xin04}.
This particle type dependence is accounted for in quark coalescence models~\cite{Fries04}. 
In these hadronization
models, hadrons are dominantly formed by coalescing massive constituent quarks from
a partonic system with the intrinsic assumption of collective flow among these partons.
These models predict a universal scaling of the observed elliptic flow $v_2$ 
and the hadron transverse momentum \pt\ with the number of constituent 
quarks $n$ (meson, $n=2$, baryon, $n=3$). 
The accordingly $n$-scaled values for $v_2$ versus \pt\ are shown in Fig.~2(d). 
Above a parton momentum \pt$/n>0.7$~\gevc, the predicted universal scaling holds within
experimental uncertainties. An exception might be $\pi$ which can be attributed
to the contribution of feeddown from resonances with $\pi$ 
in the decay channel~\cite{Xin04}. 
The successful prediction of quark coalescence models further supports the idea that
collectivity develops at the partonic stage at RHIC. 
The important question and maybe the final step to a QGP discovery at RHIC is the status
of thermalization of light quarks.

\section{Heavy-Flavor Collectivity as a probe of Light-Flavor Thermalization}
%
%\begin{figure}
%  \includegraphics[height=.3\textheight]{QuarkMasses.eps}
%  \caption{Picture to fixed height}
%\end{figure}
%
Heavy-flavor quarks are special probes because of their heavy mass. 
If chiral symmetry is restored in a QGP, light quarks obtain their small 
current masses. On the other hand, heavy quarks get almost all their mass 
from their coupling to the Higgs field~\cite{Mueller}. 
Thus, heavy quarks  stay heavy - even in a QGP. The observation 
of heavy-quark collective flow indicates multiple interactions among partons. 
This would suggest that light quarks are thermalized.
%Here, the heavy-flavor transverse elliptic flow is an especially promising early stage observable,
%while transverse radial flow might be cumulated throughout the whole collision 
%history~\cite{Sorge}.
%
\begin{figure}[hb]
  \includegraphics[height=.22\textheight]{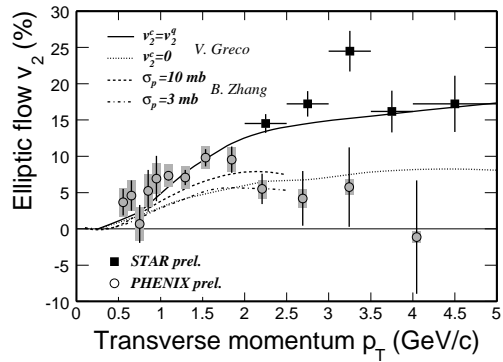}
  \caption{Results of elliptic flow measurements on electrons 
from heavy-flavor semileptonic decays~\cite{Xin_qm05} as a function of \pt\ 
from STAR~(closed squares) and PHENIX~(open circles). 
The curves show results from a quark coalescence model~\cite{Greco} assuming 
identical flow of heavy and light quarks (solid) and no heavy quark flow (dotted) 
and microscopic calculations using different partonic cross sections~\cite{Zhang} 
of 10~mb~(dashed) and 3~mb~(dash-dotted).}
\end{figure}

First results on heavy-flavor production at RHIC have been reported from
observing electrons stemming from the decay of heavy-flavor \mbox{quarks~\cite{Xin,RHIC_e}}.
Recent results of elliptic flow measurements on electrons 
from heavy-flavor semileptonic decays~\cite{Xin,Xin_qm05} 
are shown in Fig.~3 as a function of \pt\ from STAR~(closed squares) and PHENIX~(open circles). 
The electron momentum range \pt =0.5-2.0~\gevc\ corresponds to heavy-flavor hadron \pt=1.0-4.0~\gevc.
In this region, the values of $v_2$ are significantly different from zero.
The curves show results from calculations within a quark coalescence model~\cite{Greco} assuming 
identical flow of heavy and light quarks (solid) and no heavy quark flow (dotted) 
and microscopic calculations using different partonic cross sections~\cite{Zhang} 
of 10~mb~(dashed) and
3~mb~(dash-dotted). Both models support the idea of heavy-flavor collectivity at RHIC, while
the unexpectedly large cross section needed to describe the experimental data comes as a
surprise. 
%At electron momenta above 2~\gevc, other, non-collective effects become important, e.g. heavy quark
%energy loss~\cite{Kharzeev}.  
%Here, some discrepancy between both data sets exists, which is not understood yet. 
It is possible, as argued by several theorists, that elliptic flow and
energy loss of heavy quarks are correlated~\cite{theory1,theory2,theory3,theory4}.
It is therefore very interesting to study both elliptic flow and nuclear modification
factors.

However, due to the decay kinematics, 
important information on heavy-flavor dynamics is smeared \mbox{out~\cite{Kelly}}.
It seems that we do not fully understand the underlying mechanism of heavy-flavor interaction
with the dense medium. At higher \pt, therefore, it is also important to measure distributions
from directly reconstructed $D$-mesons in order to isolate the bottom contributions in collisions at RHIC.

\begin{figure}[htb]
  \includegraphics[height=.18\textheight]{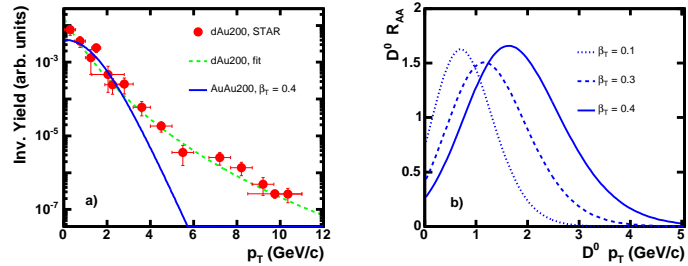}
  \caption{(a) The measured invariant yield of $D$-mesons from direct reconstruction 
through the invariant mass of decay-daughter candidates 
in \dau\ collisions at \rts=200~GeV as a function of \pt~\cite{An,An_qm04}. 
The dashed line shows the fit-result of a pQCD inspired power-law function.
A prediction from hydro-dynamically inspired model calculations is shown by the solid line.
(b) The modification of the $D^0$ spectrum 
as a function of transverse momentum for three different
flow velocities. 
}
\end{figure}
The measured invariant yield of $D$-mesons from direct reconstruction 
through the invariant mass of decay-daughter candidates 
in \dau\ collisions at \rts=200~GeV 
is shown in Fig.~4(a) as a function of \pt~\cite{An,An_qm04}.
The spectrum steeply falls with increasing momentum, followed by a long tail at high \pt.
The dashed line shows the fit-result of a pQCD inspired power-law function, 
describing the experimental data over the whole momentum range. 
A prediction from hydro-dynamically inspired model calculations is shown by the solid line.
In these calculations, a kinetic freeze-out temperature $T_{\rm fo}$ = 160 MeV and an average 
flow velocity \mbeta = 0.4 (in units of speed of light) was assumed.  
Both curves are normalized to the same yield in the momentum range \pt = 0 - 14~\gevc.
The presence of collective flow modifies the spectrum, its shape changes from concave 
(no-flow in \dau\ collisions) to convex (flow in \auau\ collisions).

The modification of the \d0 \ spectrum is further quantified by taking the ratio $R_{AA}$
of the spectra expected from \auau\ collisions (flow) relative to \dau\ collisions (non-flow).
Figure~4(b) shows the modification of the $D^0$ spectrum 
as a function of average transverse momentum for three different
flow velocities with \mbeta = 0.1~(dotted), 0.2~(dashed) and 0.4~(solid). 
The modification is in the order of 30-50\% with the maximum moving to
larger momentum with increasing flow velocity.

Due to the large multiplicities of $\pi$,K,p and the rather small production cross section
for charm-hadrons, the combinatorial background in the invariant mass distribution
is roughly 1000 times larger than the signal~\cite{Haibin}. Extending particle identification by time
of flight information will improve the statistical significance by a factor of five.
This large combinatorial background leads to systematic uncertainties of extracted charm-hadron 
yields in the order of 30\%. On the other hand, elliptic flow modulates particle yields with respect to
the reaction plane in the order of 10\%. 
To overcome these large systematic uncertainties and make precise heavy-flavor 
elliptic flow measurements feasible,
we propose to upgrade STAR with $\mu$-vertex capabilities to identify heavy-flavor hadrons through
their displaced decay vertex~\cite{HFT}.   

\section{Summary}
Elliptic flow measurements have demonstrated that partonic collectivity,
collective flow of partons, develops in 200~GeV Au+Au 
collisions at RHIC. To pin down the partonic EOS of matter produced
at RHIC, the status of thermalization in such collisions has to be addressed. 
Since the masses of heavy-flavor quarks, e.g. charm quarks, 
are much larger than the maximum possible excitation of the system 
created in the collision, heavy-flavor collective motion could be 
used to indicate the thermalization of light flavors ($u,d,s$). 
The completion of the time of flight barrel and
the proposed upgrade with a $\mu$Vertex detector for heavy-flavor identification
in STAR are well suited for achieving these goals. 
\begin{theacknowledgments}
 
Discussions with Drs. X.Dong, J.~Gonzalez, Y.~Lu, H.G.~Ritter, P.~Sorensen, 
L.~Ruan, N.~Xu, Z.~Xu and H.~Zhang are
gratefully acknowledged. 
%This work was supported under DOE contract number $N$.

\end{theacknowledgments}

%%%%%%%%%%%%%%%%%%%%%%%%%%%%%%%%%%%%%%%%%%%
%% The following lines show an example how to produce a bibliography
%% without the help of the BibTeX program. This could be used instead
%% of the above.
%%%%%%%%%%%%%%%%%%%%%%%%%%%%%%%%%%%%%%%%%%%


\begin{thebibliography}{9}

\bibitem{Karsch02} F. Karsch, Nucl. Phys. A698 (2002) 199c.

\bibitem{3_2} E.~Schnedermann, J.~Sollfrank and U.~Heinz, Phys. Rev. C48 (1993) 2462.

\bibitem{starpikp} J. Adams, {\it et al.},
Phys. Rev. Lett. 92 (2004) 112301.

\bibitem{ActaHun} K.Schweda and N.Xu, Acta Phys. Hung. A22, (2005) 103. 

\bibitem{pbm01} P. Braun-Munzinger {\it et al.},  Phys. Lett. B344 (1995) 43; 
                P. Braun-Munzinger, I. Heppe, and J. Stachel, Phys.  Lett. B465 (1999) 15. 

\bibitem{xiom130} J. Adams {\it et al.},
Phys. Rev. Lett. 92 (2004) 182301.

\bibitem{hsx98} H. van Hecke, H. Sorge, and N. Xu, 
Phys. Rev. Lett. 81 (1998) 5764.

\bibitem{Sorge} H. Sorge, Phys. Rev. Lett. 82 (1999) 2048.

\bibitem{PHENIX_v2} S.S. Adler {\it et al.}, Phys. Rev. Lett. 91 (2003) 182301.
\bibitem{STAR_v2KL} J. Adams {\it et al.}, Phys. Rev. Lett.  92 (2004) 052302.
\bibitem{STAR_v2XiO} J. Adams {\it et al.}, Phys. Rev. Lett. 95 (2005) 122301. 

\bibitem{Xin04}  X. Dong {\it et al.}, Phys.Lett. B597 (2004) 328. 

\bibitem{Fries04} For an overview see R. Fries, J. Phys. G30 (2004) 853.

\bibitem{Mueller} B. M\"uller, hep-ph/0410115.

%\bibitem{Cassing04} E.L. Bratkovskaya {\it et al.}, Phys. Rev. C71 (2005) 044901.

\bibitem{Xin} X. Dong, {\it these proceedings}.

\bibitem{RHIC_e}  K. Adcox {\it et al.}, Phys. Rev. Lett. 88 (2002) 192303; \\
J. Adams {\it et al.}, Phys. Rev. Lett. 94 (2005) 062301.

\bibitem{Xin_qm05} X. Dong, {\it Proceedings of the Quark Matter 2005 Conference},
to appear in Nucl. Phys. A. 

\bibitem{Greco} V. Greco {\it et al.}, Phys.Lett. B595 (2004) 202. 

\bibitem{Zhang} B. Zhang, L.-W. Chen, and C-M. Ko, Phys.Rev. C72 (2005) 024906.

%\bibitem{Kharzeev} Y.L. Dokshitzer and D.E. Kharzeev, Phys. Lett. B519 (2001) 199.

\bibitem{theory1}
M. Djordjevic {\it et al.}, nucl-th/0507019; Magdalena Djordjevic, Miklos Gyulassy,
and Simon Wicks, Phys. Rev. Lett. 94 (2005) 112301.

\bibitem{theory2} N. Armesto {\it et al.}, Phys. Rev. D71 (2005) 054027;
hep-ph/0501225

\bibitem{theory3} H. van Hees, V. Greco, and R. Rapp, nucl-th/0508055.

\bibitem{theory4}  G.D. Moore and D. Teaney, Phys. Rev. C71 (2005) 064904.

\bibitem{Kelly} S. Batsouli {\it et al.}, Phys. Lett. B557 (2003) 26.

\bibitem{An} A. Tai, {\it these proceedings}.

\bibitem{An_qm04} A. Tai, J. Phys. G30 (2004) S809.

\bibitem{Haibin} H. Zhang, {\it Proceedings of the Quark Matter 2005 Conference},
to appear in Nucl. Phys. A. 

\bibitem{HFT} K. Schweda, {\it Proceedings of the Quark Matter 2005 Conference},
to appear in Nucl. Phys. A. 

\end{thebibliography}
\end{document}